\documentclass[twocolumn,showpacs,aps]{revtex4}

\usepackage[dvips,final]{graphics}
\usepackage{textcomp}

\begin{document}

\title{Time-resolved homodyne characterization of individual quadrature-entangled pulses}

\author{J\'er\^ome Wenger, Alexei Ourjoumtsev, Rosa Tualle-Brouri and Philippe Grangier}

\address{Laboratoire Charles Fabry de l'Institut d'Optique, CNRS
UMR 8501, F-91403 Orsay, France.}

\begin{abstract}
We describe a simple and efficient setup to generate and
characterize femtosecond quadrature-entangled pulses. Quantum
correlations equivalent to about 2.5~dB squeezing are efficiently
and easily reached using the non-degenerate parametric
amplification of femtosecond pulses through a single-pass in a
thin (100 $\mu$m) potassium niobate crystal. The entangled pulses
are then individually sampled to characterize the non-separability
and the entropy of formation of the states. The complete
experiment is analysed in the time-domain, from the pulsed source
of quadrature entanglement to the time-resolved homodyne
detection. This particularity allows for applications in quantum
communication protocols using continuous-variable entanglement.
\end{abstract}


\maketitle

\section{Introduction}
\label{intro}

Quantum correlations  have properties which cannot be  reproduced
by the rules of classical physics \cite{EPR1935,Bell}. More
specifically, quantum entanglement is now acknowledged  as a
physical resource, which is needed to perform new quantum
information processing tasks, such as quantum teleportation, dense
coding, quantum cryptography or quantum computation
\cite{NielsenChuang}. To perform these protocols, quantum
continuous variables \cite{Braunsteinbook} have recently emerged
as a relevant alternative to discrete-levels systems. This is
particularly true in the optical domain, where entanglement of
quantum continuous variables provides a tool of major importance
for developing new quantum communication devices, based on
homodyne detection of intense beams, rather than photon-counting
detectors.

In order to characterize simply quantum communication protocols,
it is very convenient to consider them as the exchange of some
symbols, which carry the relevant information and can be accessed
individually. Hence, an important issue for quantum communication
is to develop communication schemes operating in a pulsed regime,
that are able to manipulate individually each quantum state
involved in the exchange. The analysis of such protocols is then
very easy in terms of information transfers
\cite{Shannon,CryptoCohNat}. However, since the landmark
experiment by Ou et al. \cite{KimbleOPA,KimbleOPAbis}, most of the
experiments relying on entangled quantum continuous variables have
been performed by using continuous wave (rather than pulsed)
entangled light beams
\cite{Kimbleteleport,KunchiNOPA,Schori02,BowenPRL,BowenEnt,FurusuwaDenseCod,FabreEnt1,FabreEnt2,BowenPol,JosseBis}.
In addition, the characterization of the quantum correlations was
performed in the spectral domain (i.e. by using radio-frequency
spectrum analysers), rather than in the time domain. As a
consequence, even if they are actually pulsed
\cite{LeuchsPRL,LeuchsBis}, such experiments cannot be easily used
to implement quantum communication protocols, because they do not
provide access to each individual entangled pulse. These
difficulties might be solved by developing a quantum theory of
analog modulation, evaluating the information transfer by defining
appropriate time-bandwidth limited modes, and quantizing them. But
as long as no such theory has been proposed, addressing
well-separated ``symbols" -- each associated to one quantum state
-- clearly appears to be much easier, both from a practical and a
fundamental point of view.

In this article, we present a new scheme for the generation of
quadrature-entangled light pulses, which can be individually
accessed and characterized. This experiment is based on the
non-degenerate parametric amplification of ultrafast (150~fs)
pulses through a single pass in a thin potassium niobate
(KNbO$_3$) crystal. Thanks to the high peak power of the
femtosecond pulses and the high nonlinear coefficient of the
KNbO$_3$ crystal, significant quantum correlations (equivalent to
about 2.5~dB squeezing) have been reached. The non-separability of
the entangled pulses is then directly characterized by a
time-resolved homodyne detection, which samples the quantum
properties of each individual incoming pulse.

Recently, our team has developped a similar setup which operated
in a degenerate configuration to produce pulsed squeezed states
\cite{WengerOL}. The experiment described here presents some
further developments of this scheme towards pulsed quadrature
entanglement. As said above, all our setup operates in the pulsed
regime, from the source of entanglement to the time-resolved
detection, so our analysis is completely carried out in the time
domain, and not in the frequency domain. Since the
quadrature-entangled pulses are efficiently and easily generated
through a single pass in a nonlinear crystal using non-degenerate
parametric optical amplification (NOPA), the setup provides a
simple and compact source for pulsed quadrature entanglement.

This paper is organized as follows~: first, we review some general
results concerning the manipulation and the characterization of
quadrature-entanglement (section \ref{sec:1}). Then the
experimental setup is described in details (section \ref{sec:2})
before presenting the results of our pulsed homodyne measurements
(section \ref{sec:3}).

\section{Quadrature entanglement~: principles and characterization}
\label{sec:1}

\subsection{Producing entanglement~: the non-degenerate parametric amplifier}

A well-known way to generate a two-mode state whose quadrature
components are entangled is to use an optical parametric amplifier
in a non-degenerate configuration
\cite{ReidCriterion,MillburnWalls}. Let us denote by $(X,P)$ the
quadrature components of one mode of the light field. These
quantum operators follow the commutation relation $[X,P]=2iN_0$
and thus the Heisenberg uncertainty $\Delta^2 X\,\Delta^2 P\,\geq
N_0\,^2$, where $\Delta^2(.)$ denotes the variance of the
observable ($N_0$ is a scaling constant, which equals to the
standard shot noise variance). The quadrature components
$(X_A,P_A)$ and $(X_B,P_B)$ of the output modes (``signal'' and
``idler'') of a perfect non-degenerate parametric amplifier are
given by \cite{MillburnWalls,ReidCriterion}~:
\begin{eqnarray} \label{Eq:AmpliXP}
X_{A} &=& \cosh r\,X_{A,in}+ \sinh r\,X_{B,in} \\ \nonumber P_{A}
&=& \cosh r\,P_{A,in}- \sinh r\,P_{B,in} \\ \nonumber X_{B}
&=&\cosh r\,X_{B,in}+ \sinh r\,X_{A,in} \\ \nonumber P_{B}&=&
\cosh r\,P_{B,in}- \sinh r\,P_{A,in}
\end{eqnarray}
where $(X_{A,in},P_{A,in})$ and $(X_{B,in},P_{B,in})$ stand for
the input modes and $r$ is the squeezing factor, which depends on
the strength of the nonlinear interaction. In the following, we
will consider the case where the input modes are in the vacuum
state. The state generated by NOPA is then generally called
\emph{two-mode squeezed state}, as it bears some quantum
correlations between the quadratures~:
\begin{eqnarray}
  \langle X_A\,X_B \rangle = - \langle P_A\,P_B \rangle = \sinh 2r\,N_0 \\
  \langle (X_A-X_B)^2 \rangle = \langle (P_A+P_B)^2 \rangle = 2\,e^{-2r}\,N_0
\end{eqnarray}
In the case of infinite nonlinear effects, $r\rightarrow \infty$,
the quantum correlations between the quadratures would become
perfect (e.g. $\langle (X_A-X_B)^2 \rangle \rightarrow 0$). This
case corresponds to the famous Einstein, Podolsky and Rosen 1935
\emph{Gedankenexperiment} \cite{EPR1935}.

\subsection{Characterizing entanglement~: the non-separability criterion}

For a given state composed of two modes $A$ and $B$, a relevant
question is to determine whether this state carries some
entanglement or not. For quantum continuous variables, Duan
\textit{et al} \cite{DuanSeparability} and Simon
\cite{SimonSeparability} have independently formulated a
sufficient condition for a state to be non-separable~:
\begin{equation}\label{Eq:Dscrit}
  \mathcal{I}_{\mathrm{DS}} = \frac{1}{2} \left[ \Delta^2(X_A-X_B) +
  \Delta^2(P_A+P_B) \right] < 2\,N_0
\end{equation}
Dealing with a symmetrical Gaussian two-mode state, such as the
one generated by NOPA, this condition turns out to be a necessary
and sufficient condition of non-separability. Another criterion
used to quantify the quantum correlations between the entangled
modes is the ``Reid-EPR'' criterion \cite{ReidCriterion}, which
applies to the products of the conditional variances~:
$V_{X_B|X_A}\, V_{P_B|P_A} < N_0\,^2$. It is also known that
quantum states that are non-classical according to the Reid-EPR
condition given above will automatically be non-separable
according to the Duan-Simon condition (\ref{Eq:Dscrit}), though
the reverse is not true. Thus, in the present work, we will use
the Duan-Simon criterion as the condition for non-separability.

In the case of the state generated by perfect NOPA, one gets
$\mathcal{I}_{\mathrm{DS}} = 2\,e^{-2r}\,N_0$. Thus the violation
of the Duan-Simon condition for separability occurs as soon as
$r>0$ and increases with the strength $r$ of the nonlinear
interaction. This suggests to use $\mathcal{I}_{\mathrm{DS}}$ to
quantify the entanglement. However, a relevant measure of
entanglement has to satisfy some constraints given in
\cite{Plenio,EisertEnt}. Unfortunately,
$\mathcal{I}_{\mathrm{DS}}$ does not fulfil these constraints and
thus does not provide a ``true'' measure of entanglement.
Nevertheless, $\mathcal{I}_{\mathrm{DS}}$ is a direct witness of
non-separability and can be used to compute a relevant measure of
entanglement for Gaussian states, as we will see now.

\subsection{Quantifying entanglement~: the entropy of formation}

For a bipartite pure state, the amount of entanglement is uniquely
defined by the Von Neumann entropy of one of the sub-state
\cite{EisertEnt}. However, for a mixed state, the definition of
the quantity of entanglement is not unique, and has led to
different theoretical proposals. Among them, let us pay some
attention to the entropy of formation, introduced by Wooters in
\cite{WootersEoF}. This quantity represents the number of qubit
pairs necessary to prepare the observed correlations. An
interesting point of this quantity is that it has been explicitly
calculated in the case of Gaussian states by Giedke and coworkers
\cite{GiedkeEoF}. For a symmetric two-mode Gaussian state whose
covariance matrix reads
\begin{equation} \label{Eq:CovMat1}
\gamma= \left(
\begin{array}{cccc}
V & 0 & K_x & 0 \\[2mm]
0 & V & 0 & -K_p \\[2mm]
K_x & 0 & V & 0 \\[2mm]
0 & -K_p & 0 & V
\end{array} \right)
\end{equation}
the entropy of formation is then given by the formula~:
\begin{equation}
 E_{\mathrm{F}} = f \left(\,\sqrt{(V-K_x)(V-K_p)}\,\right)
\end{equation}
with $f(x) = c_+(x)\,\log_2 c_+(x) - c_-(x)\,\log_2 c_-(x)$ and
$c_{\pm}(x)=[x^{-1/2} \pm x^{1/2}]^2/4$. Consequently, for a state
generated by NOPA, the entropy of formation is directly linked to
the Duan-Simon quantity $\mathcal{I}_{\mathrm{DS}}$ as noticed in
\cite{JosseBis}~:
\begin{equation}\label{Eq:EoF}
 E_{\mathrm{F}} = f \left(\,\frac{\mathcal{I}_{\mathrm{DS}}}{2\,N_0}\,\right)
\end{equation}

\subsection{Simple experimental measurements of quadrature
entanglement} \label{SSec:caract}

\begin{figure}[t]
\center \includegraphics{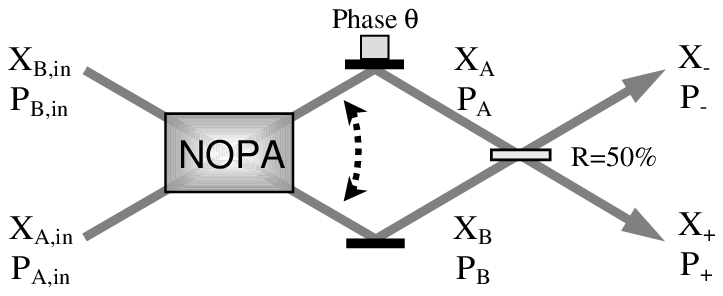}
\caption{Simplified scheme of the experimental setup used to
characterize the quadrature entanglement produced during
single-pass non-degenerate parametric amplification (NOPA). The
beam $(X_+,P_+)$ is then measured by a time-resolved homodyne
detection.} \label{Fig:SchemPPe}
\end{figure}

The method experimentally developed here to characterize the
Gaussian quadrature-entangled states is largely inspired from
these theoretical results and aims at directly measuring the
Duan-Simon quantity $\mathcal{I}_{\mathrm{DS}}$. The principle of
our scheme is depicted on Fig.~\ref{Fig:SchemPPe}. The two
entangled modes generated by NOPA are mixed on a beamsplitter of
$R=50\%$ reflectivity, while one output mode of the beamsplitter
is sent to a time-resolved homodyne detection for pulse-sensitive
quadrature measurements.

Let $\theta$ denote the relative phase between the entangled
pulses. When the two entangled modes $A$,$B$ are mixed in phase on
the beamsplitter, $\theta=0$, the output mode of the beamsplitter
$(X_{+},P_{+})$ reads~:
\begin{eqnarray}
  X_{+} &=& \frac{1}{\sqrt{2}}\left( X_A + X_B \right) = \frac{e^r}{\sqrt{2}}\left( X_{A,in} + X_{B,in}
  \right)\\ \nonumber
P_{+} &=& \frac{1}{\sqrt{2}}\left( P_A + P_B \right) =
\frac{e^{-r}}{\sqrt{2}}\left( P_{A,in} + P_{B,in}
  \right)
\end{eqnarray}
Thus the state at the output of the beamplitter is a squeezed
state, and the corresponding variances in quadratures are either
above or below the shot noise level~: $\Delta^2 X_{+} =
e^{2r}\,N_0$, $\Delta^2 P_{+} = e^{-2r}\,N_0$. Measuring $\Delta^2
P_{+} = \Delta^2(P_A+P_B)/2$ with the homodyne detection gives
directly one term to be used in the Duan-Simon quantity
$\mathcal{I}_{\mathrm{DS}}$ (\ref{Eq:Dscrit}).

If a relative phase of $\theta=\pi$ is set between the two
entangled modes $A$,$B$, the output mode of the beamsplitter
$(X_{+},P_{+})$ then reads~:
\begin{eqnarray}
  X_{+} &=& \frac{1}{\sqrt{2}}\left( X_A - X_B \right) = \frac{e^{-r}}{\sqrt{2}}\left( X_{A,in} - X_{B,in}
  \right)\\ \nonumber
P_{+} &=& \frac{1}{\sqrt{2}}\left( P_A - P_B \right) =
\frac{e^{r}}{\sqrt{2}}\left( P_{A,in} - P_{B,in} \right)
\end{eqnarray}
This is still a squeezed state, but the squeezed quadrature has
now turned to the $X$ quadrature. Measuring $\Delta^2 X_{+} =
\Delta^2(X_A-X_B)/2$ with the homodyne detection, one gets
directly the other term of the Duan-Simon quantity
$\mathcal{I}_{\mathrm{DS}}$. Thus by mixing the entangled pulses
on a beamsplitter and controlling the relative phase $\theta$
between them and between the local oscillator used in the homodyne
detection, one can measure all the relevant parameters to express
the Duan-Simon quantity $\mathcal{I}_{\mathrm{DS}}$. More
generally, one can easily show that for any phase $\theta$ the
quadrature $\sin(\frac{\theta}{2})\,X_{+}+
\cos(\frac{\theta}{2})\,P_{+}$ is squeezed while the quadrature
$\cos(\frac{\theta}{2})\,X_{+}-\sin(\frac{\theta}{2})\,P_{+}$ is
anti-squeezed. Thus dephasing the EPR beams by $\theta$ before the
recombination makes the phase-space uncertainty ellipse turn by an
angle of $\theta/2$, but it keeps the ellipticity constant.

One interesting feature of this characterization method is that it
provides simply the necessary quantity $\mathcal{I}_{\mathrm{DS}}$
to check for the non-separability from Eq.(\ref{Eq:Dscrit}) and to
quantify the entanglement from Eq.(\ref{Eq:EoF}). Only one
homodyne detection setup is needed, which greatly simplifies the
experiment. This method appears also quite similar to the setups
used for dense coding experiments
\cite{BKdensecod,Ralphdensecod,KunchiNOPA}. Let us now turn to the
physical implementation of these principles in the pulsed regime.

\section{Experimental set-up}
\label{sec:2}

We have recently developed a new scheme for pulsed squeezed light
generation \cite{WengerOL}. This experiment is based on the
degenerate parametric amplification of femtosecond pulses. Here we
extend this experiment and pass to the non-degenerate regime of
parametric amplification.

\begin{figure}[t]
\center \includegraphics{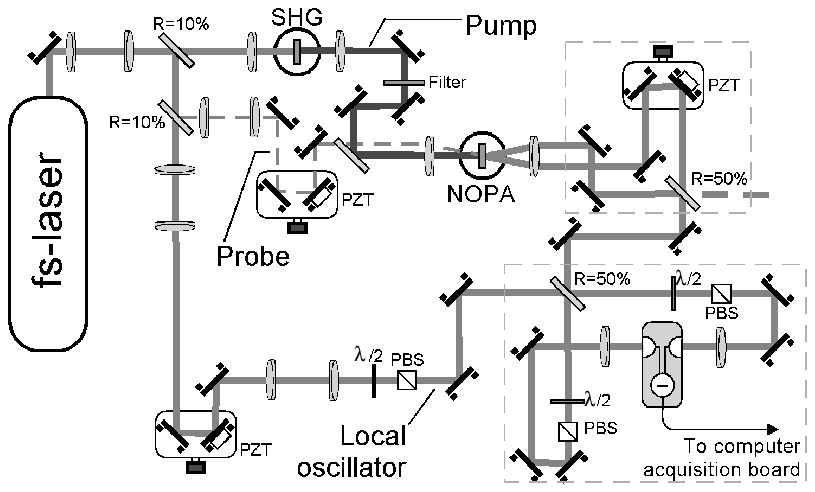} \caption{Experimental
setup. SHG second harmonic generation; NOPA non-degenerate
parametric amplification; PZT piezoelectric transducer; PBS
polarizing beamsplitter.} \label{Fig:Expsetup}
\end{figure}

The experimental scheme is presented on Fig.~\ref{Fig:Expsetup}.
The initial pulses are obtained from a cavity-dumped
titanium-sapphire laser (Tiger-CD, Time-Bandwidth Products),
delivering nearly Fourier-transform limited pulses at 846~nm, with
a duration of 150~fs, an energy of 40~nJ, and a pulse repetition
rate of 780 kHz. These pulses are frequency doubled in a single
pass through a thin (100~$\mu$m) crystal of potassium niobate
(KNbO$_3$). This crystal is set inside a small vacuum chamber and
peltier-cooled down to about $-14^{\circ}$C to obtain non-critical
(90 degrees) type-I phase-matching for second harmonic generation
(SHG) at 846 nm. The choice for a thin crystal length allows for a
wide phase-matching bandwidth and avoids the conditions of large
group-velocity mismatch, contrary to the previously reported use
of thick KNbO$_3$ crystals \cite{Weiner,Xiao99}. Even for the
short interaction length used here, KNbO$_3$ proved to be suitable
to our applications thanks to its high non linear coefficient
(about 12 pm/V) and non-critical phase-matching. Typically, the
SHG efficiency obtained is about $28\%$ (corrected from losses).

A small fraction ($1\%$) of the fundamental beam is taken out to
serve as a probe to study classical parametric amplification
occurring in a similar KNbO$_3$ crystal used in a single-pass
type-I spectrally degenerate but spatially non-degenerate
configuration. This spatial non-degeneracy is obtained by shifting
horizontally the probe beam before focussing inside the crystal.
Thanks to the focussing lens, this shift is transformed into an
angular difference between the propagation directions of the pump
and the probe beam within the crystal (see
Fig.~\ref{Fig:Expsetup}). For small angular differences of about
3$^{\circ}$ within the crystal, the phase-matching condition can
still be fulfilled by lowering the temperature of the crystal by a
few degrees. The average power of the probe beam after the
nonlinear interaction is directly measured by a standard Silicium
photodiode, which allows to estimate for the parametric gain. For
a well-chosen angular shift so that the probe and the pump wave
vectors do not overlap, the parametric gain does not depend of the
relative phase between the pump and the probe beam. This condition
indicates that the non-degenerate configuration is strictly
reached other the whole spatial extend of the beams. We have
experimentally optimised the overlap between the pump and the
probe beam to maximise the amplification gain. For the best
configuration, the probe waist was set to be about $\sqrt{2}$
times smaller than the pump waist inside the crystal. Then, the
best amplification obtained was 1.24\,. This gain was checked to
be independent of the probe phase and average power.

The two beams (``signal'' and ``idler'') resulting from the
classical parametric amplification of the probe
are mixed on a beamsplitter of $R=50\%$ reflectivity, in order to
implement the entanglement characterization method described in
section \ref{SSec:caract}. One output mode of the beamsplitter is
then sent to the balanced homodyne detection, and interferes with
the local oscillator beam (LO). The homodyne detection is set to
be directly sensitive to the incoming pulse distribution in the
time domain. For each pulse, the fast acquisition board (National
Instruments PCI-6111E) samples one value of the signal quadrature
in phase with the local oscillator
\cite{Smithey,HansenOL,CryptoCohNat}. The data presented below are
obtained directly from these individual pulse measurements. Pulsed
homodyning is technically more challenging than frequency-resolved
homodyning because low-frequency noises cannot be filtered out.
Each arm of the detection has to be carefully balanced (with a
typical rejection better than $10^{-4}$) even for ultra-low
frequency noises. By blocking the squeezed beam, the detection was
checked to be shot-noise limited in the time domain, showing a
linear dependence between LO power and the noise variance up to
$2.5~10^8$ photons per pulse at a repetition rate of 780 kHz and
in the femtosecond regime. The electronic noise was low enough to
ensure a ratio larger than 11 dB between shot noise and electronic
noise variances.

\section{Pulsed homodyne measurements}
\label{sec:3}

\subsection{Characterization of the non-separability criterion}

\begin{figure}[!t]
\center \includegraphics{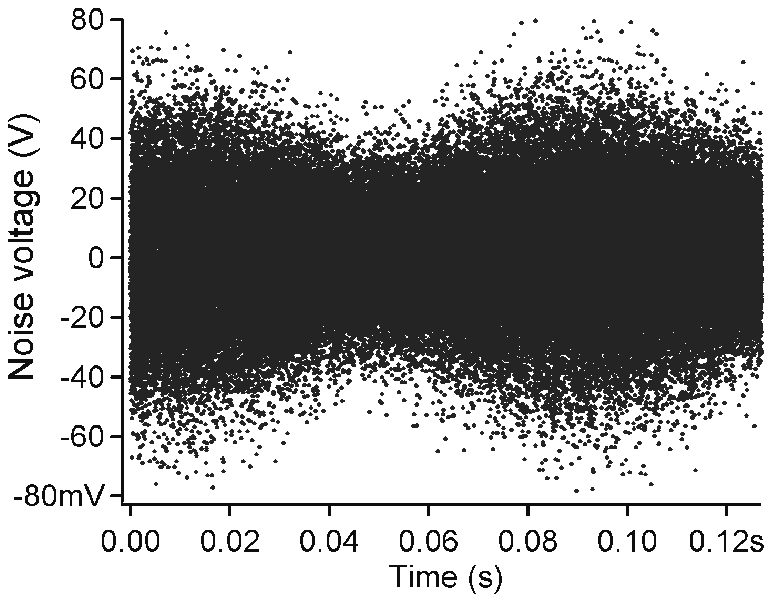} \caption{Recorded
noise pulses while linearly scanning the LO phase, when the
entangled beams are mixed in phase ($\theta=0$) at the $50-50$
beamsplitter. Each dot corresponds to the measurement of one
incoming pulse.} \label{Fig:squeez1}
\end{figure}

\begin{figure}[!t]
\center \includegraphics{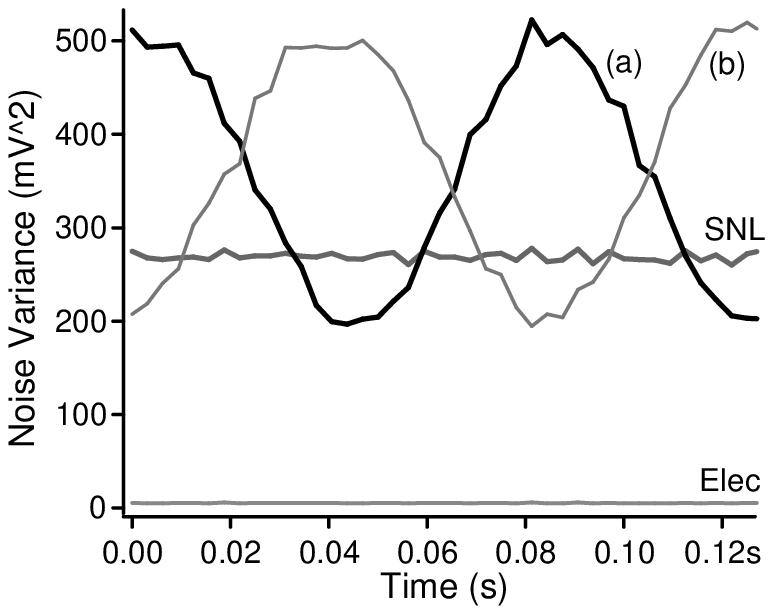} \caption{Quadrature
noise variance of the recombined beam $(X_+,P_+)$ (plotted in a
linear scale and computed over blocks of 2,500 samples) while
linearly scanning the LO phase, together with the shot noise level
(SNL) and the electronic noise level. The curve (a) corresponds to
the case when the entangled beams are mixed in phase ($\theta=0$),
while they are dephased by $\theta=\pi$ for the curve (b).}
\label{Fig:squeez2}
\end{figure}

The probe beam being blocked, the amplifier operates in a pulsed
parametric down-conversion regime to generate a two-mode squeezed
state. A first piezo-electric transducer allows to finely control
the relative phase $\theta$ between the two entangled pulses
before the recombination beamsplitter. A second piezo transducer
is also used to control the relative phase of the local
oscillator, which commands the quadrature component measured by
the homodyne detection. Fig.~\ref{Fig:squeez1} displays the
recorded noise pulses while scanning the local oscillator phase,
when the quadrature-entangled beams are mixed in phase,
$\theta=0$. Fig.~\ref{Fig:squeez2} presents the corresponding
quadrature variance while scanning the local oscillator phase,
when the entangled beams are in phase, $\theta=0$ (a), or dephased
by $\theta=\pi$ (b) at the $50-50$ beamsplitter. As expected for
the squeezed states generated by recombining the entangled pulses,
the measured noise variance passes below the shot noise level
(SNL) for some phase values of the local oscillator. The measured
correlation variance (with no correction) is $0.70\,N_0$ (-1.55~dB
below the shot noise level SNL), while the corresponding
anti-squeezed variance is $1.96\,N_0$ (2.92~dB above SNL).

To characterize the entanglement of the two-mode state produced by
NOPA, it appears logical to compensate for the homodyne detection
efficiency. The procedure to measure this detection efficiency is
well established from squeezing experiments \cite{WengerOL}. The
overall detection efficiency is given as $\eta = \eta_T \eta_H^2
\eta_D = 68\%$, where we  have independantly measured the overall
transmission $\eta_T =93\%$, the mode-matching visibility $\eta_H
= 88\%$ (obtained from the interference fringes between the local
oscillator and a seed beam set for maximum classical
amplification), and the detectors efficiency (Hamamatsu S3883)
$\eta_D = 94.5\%$. Given this evaluation of the homodyne
efficiency, one can evaluate the correlation variance of the
pulses before the homodyne detection, which is found to be
$0.56\,N_0$ (-2.52~dB below SNL).

This experiment was repeated several times for different relative
phases $\theta$ between the entangled pulses before the
recombination. This allowed us to check the symmetry of the state
produced~: we measured the same correlation variance for
$\theta=0$, $\theta=\pi$ and several other phases (within
reasonable statistical errors less than $0.01\,N_0$, see
Fig.\ref{Fig:squeez2}). Consequently, we can state that the
correlations between the quadratures are equal to
$\Delta^2(X_A-X_B)/2 = \Delta^2(P_A+P_B)/2 = 0.56\,N_0$. The
Duan-Simon quantity amounts then to $\mathcal{I}_{\mathrm{DS}} =
1.12\,N_0$, which is clearly below the threshold for separability
($2\,N_0$) given by Eq.~(\ref{Eq:Dscrit}). This results attests
for the quantum entanglement (non-separability) of the pulsed
state generated by NOPA. One may also quantify the entanglement
available thanks to the entropy of formation given by
Eq.~(\ref{Eq:EoF}), which provides $E_{\mathrm{F}}=0.44$~ebit.

\subsection{Characterization of the covariance matrix}

Theoretically, a general Gaussian state is fully characterized by
its mean values of quadratures together with its covariance matrix
$\gamma$, which comprises the second moments of the conjugate
quadratures $X,P$. For a two-mode state, the general covariance
matrix $\gamma$ contains 16 terms. In our case, the two-mode
squeezed state emerging from NOPA is generated and handled in a
symmetric way. This point was experimentally checked as discussed
in the previous subsection. Consequently, the covariance matrix
for our state may be reduced to the form given by
Eq.~(\ref{Eq:CovMat1}), with $V\,N_0=\Delta^2 X_A = \Delta^2 P_A
=\Delta^2 X_B =\Delta^2 P_B$ and $K_x\,N_0 = K_p\,N_0 =
\frac{1}{2}\langle X_A X_B + X_B X_A \rangle = -\frac{1}{2}\langle
P_A P_B + P_B P_A \rangle$. This means that for the optimal choice
of quadratures there is no cross-quadrature correlations. Let us
also point out that for Gaussian states the covariance matrix can
always be reduced to this form using local unitary operations
\cite{DuanSeparability}.

\begin{figure}[!t]
\center \includegraphics{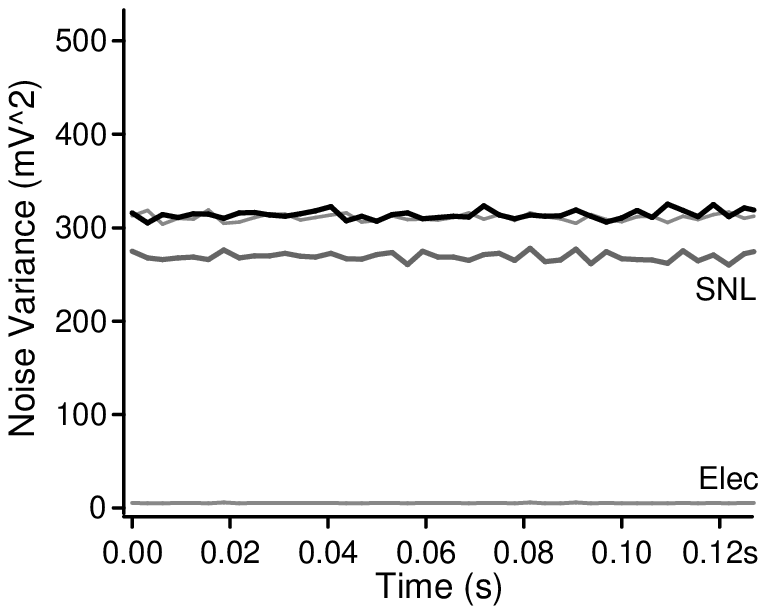} \caption{Quadrature
noise variance (dark curve) of one entangled beam (plotted in a
linear scale and computed over blocks of 2,500 samples) while
linearly scanning the LO phase. The other entangled beam is
blocked before the $50-50$ beamsplitter. The second light gray
curve above the SNL corresponds to the reverse situation by
blocking the other beam.} \label{Fig:squeez3}
\end{figure}

By blocking one entangled beam before the $50-50$ beamsplitter, we
have measured the diagonal terms of the covariance matrix (up to
the transmission of $50\%$ of the beamsplitter and the homodyne
detection efficiency $\eta$). Our results using the time-resolved
homodyne detection are displayed on Fig.~\ref{Fig:squeez3}. For
the two entangled beams, we have measured a variance of $1.17 \pm
0.01\,N_0$, which can be related to $V= 1.50$ before the $50-50$
beamsplitter (to get this value, we have corrected for the $50\%$
transmission of the beamsplitter and the homodyne detection
efficiency $\eta=68\%$). The off-diagonal terms in the covariance
matrix (\ref{Eq:CovMat1}) can be evaluated by noticing that thanks
to our squeezing measurement $\Delta^2 X_+$ we get~:
\begin{eqnarray}
  \Delta^2 X_+ &=& \frac{1}{2}\, \Delta^2(X_A-X_B) \\ \nonumber
  &=& \frac{1}{2}\,\left(\Delta^2 X_A + \Delta^2 X_B - \langle X_A X_B +
X_B X_A \rangle
  \right)\\ \nonumber
  &=& (V - K_x)\,N_0
\end{eqnarray}
With $V=1.50$ and $\Delta^2 X_+ /N_0=0.56$, we get directly
$K_x=K_p=0.94$. Thus the reconstructed covariance matrix of the
experimental entangled pulses reads~:
\begin{equation} \label{Eq:CovMat2}
\gamma= \left(
\begin{array}{cccc}
1.50 & (0) & 0.94 & (0) \\[2mm]
(0) & 1.50 & (0) & -0.94 \\[2mm]
0.94 & (0) & 1.50 & (0) \\[2mm]
(0) & -0.94 & (0) & 1.50
\end{array} \right)
\end{equation}
The values set to zero due to symmetry considerations are
indicated by parenthesis. A similar procedure has already been
followed by Bowen \emph{et al} \cite{BowenEnt}, but in the case of
continuous-wave light detected around 3.5 or 6.5~MHz using a
spectrum analyser.

\section{Conclusion}
\label{sec:4}

An efficient setup to generate and characterize femtosecond
quadrature-entangled pulses is presented, using a complete time
domain analysis. Pulsed quadrature-entangled beams are efficiently
generated using non-degenerate parametric amplification in a
single-pass configuration. This setup leads to correlations
equivalent to about 2.5~dB squeezing, that have been recorded
using a time-resolved homodyne detection. The non-separability of
the states has then been characterized using the Duan-Simon
criterion with $\mathcal{I}_{\mathrm{DS}} = 1.12\,N_0$ ($<
2\,N_0$), and the available entanglement has been quantified using
the entropy of formation to $E_{\mathrm{F}}=0.44$~ebit. The
association of this simple and compact source for quadrature
entanglement together with the time-resolved homodyne detection
provides all the ressources for future quantum communication
protocols using continuous variable entanglement.

\end{document}